\documentclass[a4paper]{jpconf}
\usepackage{graphicx}
\usepackage{amssymb}
\usepackage{amsmath}
\usepackage{color,soul}
\usepackage{wrapfig}

\begin{document}
  \title{Glitches in neutron stars with magnetically decoupled core}
  \author{O A Goglichidze and D P Barsukov}
  \address{Ioffe Institute, st. Polytekhnicheskaya, 26, Saint-Petersburg, 194021, Russian Federation.}
   \ead{goglichidze@gmail.com}
  \begin{abstract}
    The magnetically decoupled core model was proposed earlier as a way to solve the problem of  inconsistency between the neutron star long-period precession and superfluid vortex pinning which is the base of most theories of pulsar glitches. 
    It was assumed that the pinning takes place in the region of the neutron star core which, being magnetically decoupled, can rotate relative to the crust. 
    In present work some aspects of the glitch spin-up stage in the framework of the proposed model are discussed.     
    Estimated spin-up time-scales are compared with observational data. According to the estimations the Crab pulsar is likely posses a magnetically decoupling core region, while the results for the Vela pulsar is more controversial.

  \end{abstract}
  \section{Introduction}
    There are several indications that isolated neutron stars can precess with long periods. 
    In 2000, Stairs et al.  \cite{StairsLyneShemar2000} reported discovery of correlated periodic variations in spin-down rate and beam shape of PSR B1821-11. 
    The most favorable explanation to this phenomenon is the precession of the neutron star with period $T_p \approx 500$ days \cite{Jones2012,AshtonJonesPrix2016} (see, however, Stairs et al. 2019 \cite{StairsEtAl2019}).
    Several pulsars show periodic variations in  spin-down rate without significant correlation with beam shape \cite{KerrHobbsJohnstonShannon2016}. The time-scales of the variations are  0.5 - 1.5~yr. 
    
   Besides the directly observed variations, there are the observational data which can also be interpreted as manifestations of neutron star precession but with much larger periods.
  Lyne et al. \cite{LyneEtAl2013} found a steady increase in separation between the main pulse and the interpulse of the Crab pulsar at 0.62$^\circ \pm 0.03^\circ$ per century.
  They concluded that this is a consequence of the increase of the pulsar inclination angle.
  The rate of the increase seems to be too large for secular  evolution but it can be ensured by the free precession with period $\sim 10^2$ yr \cite{ArzamasskiyEtAl2015}.
  Biryukov et al. \cite{BiryukovBeskinKarpov2012} argued that anomalously large braking indices indicate that the pulsar spin-down rate can oscillate at the time-scale of $10^3-10^4$ yr. This oscillation can be caused by the precession with corresponding periods.
   The stellar magnetic field by itself should make neutron star precess at such time-scale \cite{Melatos2000}.
   
    However, the long-period ($\gtrsim $1 yr) precession is difficult to coexist with neutron vortex pinning required by most pulsar glitch theories \cite{HaskellMelatos2015}. 
    On the one hand, the glitch theories require at least about 1 percent of total stellar moment of inertia to be contained in pinned superfluid ($I_s/I \sim 10^{-2}$) \cite{AnderssonGlampedakisHoEspinoza2012}.
    On the other hand, the pinned superfluid makes a pulsar precess with angular frequency \cite{Shaham1977} 
    \begin{equation}
      \label{eq:omega_p_shaham}
      \omega_p \approx (I_s/I) \Omega \sim 10^{-2} \Omega
    \end{equation}
    which is many order of magnitude larger than the observed values.
    Several attempts to attack this problem with more detailed models of pinned vortices dynamics \cite{SedrakianWassermanCordes1999,LinkCutler2002,Alphar2005}
  or with assuming that the pinning is absent  \cite{Link2006,KitiashviliGusev2008}
  were made. 
  However, to date the problem remains current~\cite{JonesAshtonPrix2017}.

  \section{Precession of a neutron star with magnetically decoupled core}
  \label{sec:precession}
    A possible way to solve the problem was proposed by Barsukov et al. \cite{KTBY2017,BGKV2017,GB2019} who assumed that in some neutron stars at least a part of the core, being magnetically decoupled from the crust, can maintain the rotation with slightly different angular velocity.  In this case, if the neutron vortex pining takes place in that core region, it almost does not affect the observed neutron star precession.  Qualitatively, it could be shown with the following simple model.
    Let us assume that a neutron star consists of three dynamically distinguished components namely the c- (crust), the g- (glitching) and the r- (recovery) component. 
    Here, the c-component is the external component whose rotation is directly observed, while the g- and the r-component are the internal core components.
    One can write 
      \begin{equation}
    \label{eq:dMi_three_comps}
    d_t\vec{M}_i = \sum_{j \neq i} \vec{N}_{ji} + \vec{K} \delta_{ic},
  \end{equation}
  where $i, j = c, g, r$, $\vec{M}_i$ is the angular momentum of the $i$-th component, $\vec{K}$ is the external electromagnetic torque, symbol $\delta_{ic}$ means that the external torque acts on the c-component,
  $\vec{N}_{ij}$ is the torque acting on the $i$-th component from the $j$-th component.
  
  For the sake of simplicity all the components are supposed to rotate as rigid bodies with angular velocities $\vec{\Omega}_i$.
  The corresponding angular momenta  can be represented in the following form: \\
  \begin{subequations}
    \begin{minipage}{0.34\textwidth}
      \begin{equation}
        \label{eq:Mc_three_comps}
        \vec{M}_c = I_c \vec{\Omega}_c + I_c \epsilon_c (\vec{\Omega}_c\cdot\vec{e}_c)\vec{e}_c,
      \end{equation}
    \end{minipage} \ \ \
    \begin{minipage}{0.42\textwidth}
      \begin{equation}
        \label{eq:Mg_three_comps}
        \vec{M}_{g} = I_g\vec{\Omega}_g + I_g \epsilon_g (\vec{\Omega}_g\cdot\vec{e}_g)\vec{e}_g + {L}_\mathrm{sf}\vec{e}_g,
      \end{equation}
    \end{minipage} \ \ \
    \begin{minipage}{0.2\textwidth}
      \begin{equation}
       \label{eq:Mr_three_comps}
       \vec{M}_r = {I}_r \vec{\Omega}_r.
      \end{equation}    
    \end{minipage} 
  \end{subequations}
  \\ \\  \vspace{-0.2cm} \\ 
  Here, $I_i$ are the component moments of inertia, $\vec{e}_c$ and $\vec{e}_g$ are the symmetry axes of the  c- and the g-component,   $\epsilon_c$ and $\epsilon_g$ are the corresponding oblateness parameters.
  The g-component, besides the matter rotating with angular velocity $\vec{\Omega}_g$, is supposed to contain a pinned superfluid with angular momentum $L_\mathrm{sf}$. For simplicity, we assume that the pinned superfluid angular momentum is directed along $\vec{e}_g$.
  The r-component is assumed to be spherically symmetric. 
    
    It is supposed that the interaction torques can be represented in the following form:
    \begin{equation}
      \label{eq:N_{ij}}
      \vec{N}_{ij} = I_j\alpha_{ij}(\vec{\nu}_{ij}\cdot\vec{e}_\Omega)\cdot{e}_\Omega + I_j \beta_{ij} [\vec{e}_\Omega\times[\vec{\nu}_{ij}\times\vec{e}_\Omega]]+ I_j \gamma_{ij} [\vec{e}_\Omega\times\ \vec{\nu}_{ij}],
    \end{equation}
    where $\vec{\nu}_{ij} = \vec{\Omega}_i - \vec{\Omega}_j$, $|{\nu}_{ij}| \ll \Omega_c$,  and $\{\alpha_{ij}, \beta_{ij}, \gamma_{ij}\} = (I_i/I_j)\{\alpha_{ji}, \beta_{ji}, \gamma_{ji}\}$ are formally introduced interaction coefficients.
    
    Speaking in these terms, it is usually assumed that the c- and the g-component are rigidly coupled and they can be considered as a single c-component containing pinned superfluid. 
    Hence, even if the superfluid is pinned in the NS core, it is practically a part of the ``crust'' component. 
    This is a good approximation if the whole NS core is penetrated by magnetic field lines coupling it with the crust at the very short time-scales \cite{Easson1979b}.
    However, a possible NS core region formed by closed magnetic field lines can rotate with angular velocity $\vec{\Omega}_g$ different from $\vec{\Omega}_c$ \cite{GlampedakisLasky2015}. It could be a region occupied by strong toroidal magnetic field or by the tangles of closed flux tubes which could be formed from chaotic small-scale magnetic field.
    Barsukov et al. have considered an opposite configuration of  weak gc-ineterqction (i.e. $\alpha_{gc}, \beta_{gc}, \gamma_{gc}  \ll \Omega_c$).
    In this case, there is a long lived slow precession mode with angular frequency (see formula (106) in the paper of Goglichidze and Barsukov 2019 \cite{GB2019})
%
%
    \begin{equation}
      \label{eq:prec_mode_c_three_comps}
      \omega_p \approx  \epsilon_c \Omega_c - \frac{\epsilon_c}{1+\epsilon_c}\left(\gamma_{gc} + \gamma_{rc}- i\beta_{gc} -i \beta_{rc}\right).
    \end{equation}
    This expression, in contrast to formula  \eqref{eq:omega_p_shaham},  does not contain $I_s$. Therefore, the oscillations can be slow due to smallness of the c-component oblateness parameter. 
    
    The critically important point of the model is the assumption that the precessing neutron star does not contain substantial amount of pinned superfluid in it's crust. It could be due to smallness of crust pinning force itself \cite{Jones1991} or the vortices are unpinned by the Magnus force caused by the precession \cite{LinkCutler2002}.

    \section{Glitches in magnetically decoupled core}
    \label{eq:glitches}
    In the linear in $\vec{\nu}_{ij}$ approximation, system of equations \eqref{eq:dMi_three_comps} splits into two independent subsystems describing ``perpendicular'' and ``parallel'' departures from the solid-body rotation \cite{GB2019}.
    Strictly speaking, during a glitch both ``parallel'' and ``perpendicular'' modes could be excited. 
    However, such small ``perpendicular'' excitations  can hardly be observed. We will not consider the ``perpendicular'' excitations in present paper.
    This allows us to use absolute values instead of vectors.
    
    Studying glitch dynamics, one can not stay in the perfect pinning approximation. Therefore, the g-component superfluid should be treated as a distinguish dynamical component. Let us put 
    \begin{equation}
      \label{eq:L_sf}
      L_\mathrm{sf} = I_s \Omega_s, 
    \end{equation} 
    where 
    $\Omega_s$ is the superfluid macroscopic angular velocity. 
    {This is a rough approximation to consider the superfluid as a rigid body rotator but it allows us to obtain some preliminary estimations before  more detailed hydrodynamical simulations will done.
    
    The superfluid is assumed to interact only with the g-component.  Between glitches the interaction torque is assumed to have the following form \cite{Link2014}:
    \begin{equation}
      \label{eq:N_gs}
      N_{gs} = I_s \alpha_{gs}\nu_{gs} \exp\left(-\frac{A(\nu_{gs})}{kT}\right),
    \end{equation}
    where $A(\nu_{gs})$ is the activation energy for unpinning, $kT$ is the temperature in the g-component in energy units,  $\alpha_{gs}$ is the coefficient describing the gs-interaction in the absence of the pinning.
    
    Let us represent the external torque as ${K} = - I |\dot{\Omega}|$, where $|\dot{\Omega}|$ is the rate of neutron star steady-state spin-down. This expression together with  expressions \eqref{eq:Mc_three_comps}-\eqref{eq:Mr_three_comps}, \eqref{eq:N_{ij}}, \eqref{eq:L_sf} and \eqref{eq:N_gs} can be substituted into equations \eqref{eq:dMi_three_comps}. 
    Representing $\nu_{ij} = \nu_{ij}^{eq} + \delta \nu_{ij}^{eq}$, where  $\nu_{ij}^{eq}$ are the steady-state departures caused by $|\dot{\Omega}|$ and $\delta \nu_{ij}$ are the corrections  excited during the glitch, we can express:
    \\ 
    \begin{subequations}
     \begin{minipage}{0.56\textwidth}
  \begin{equation}
    \label{eq:nu_gc_eq}
    \nu_{gc}^{eq} = |\dot{\Omega}|\frac{\alpha_{cr}+\alpha_{gr}+\alpha_{rg}+(I_s/I_g)(\alpha_{cr} + \alpha_{gr})}{\alpha_{cg}\alpha_{rc}+\alpha_{cg}\alpha_{gr}+\alpha_{cr}\alpha_{rg}},
  \end{equation}     
     \end{minipage}   
      \begin{minipage}{0.44\textwidth}
   \begin{equation}
    \label{eq:nu_gr_eq}
    \nu_{gr}^{eq} = |\dot{\Omega}|\frac{\alpha_{cr}-\alpha_{cg}+(I_s/I_g)\alpha_{cr}}{\alpha_{cg}\alpha_{rc}+\alpha_{cg}\alpha_{gr}+\alpha_{cr}\alpha_{rg}},
  \end{equation}    
     \end{minipage}   
     \end{subequations}   \ \\ 
    and $\nu_{gs}^{eq}$ can be found as a root of transcendent equation
    \begin{equation}
      \label{eq:nu_gs_eq}
      \nu_{gs}^{eq} \exp\left(-\frac{A(\nu_{gs}^{eq})}{kT}\right) = - \frac{|\dot{\Omega}|}{\alpha_{gs}}.
    \end{equation} 
   In practice, $\nu_{gs}^{eq}$ is very close to the critical lag for unpinning $\omega_\mathrm{crit}$ \cite{Link2014}.
    For  corrections $\delta \nu_{ij}$ the following equations could be obtained
    {\small
    \begin{align}
      \label{eq:d_dnu_gc} 
      d_t\delta \nu_{gc} &= -\left[ \frac{1}{\tau_{gc}}+ \frac{I_r}{I_r+I_g}\frac{1}{\tau_{gr}}\right]\delta\nu_{gc} - \left[ \frac{I_r}{I_r+I_c}\frac{1}{\tau_{rc}}- \frac{I_r}{I_r+I_g}\frac{1}{\tau_{gr}}\right] \delta\nu_{rc}- \\  
      & -   \frac{I_s}{I_g}|\dot{\Omega}|\left[ 1-\left(1+\frac{\delta\nu_{gs}}{\nu_{gs}^{eq}}\right) e^{-b \delta\nu_{gs}}\right], \nonumber
    \end{align}
    \begin{align}
      \label{eq:d_dnu_gs}
      d_t\delta \nu_{gs} &= -\left[ \frac{I_c}{I_g+I_c}\frac{1}{\tau_{gc}} + \frac{I_r}{I_r+I_g}\frac{1}{\tau_{gr}}\right] \delta\nu_{gc} + \frac{I_r}{I_r+I_g}\frac{1}{\tau_{gr}}  \delta\nu_{rc} - \\
      & -\frac{I_s+I_g}{I_g}  |\dot{\Omega}|\left[ 1-\left(1+\frac{\delta\nu_{gs}}{\nu_{gs}^{eq}}\right)e^{-b \delta\nu_{gs}}\right], \nonumber
    \end{align}
    \begin{align}
      d_t\delta \nu_{rc} &=  -\left[\frac{1}{\tau_{gc}} - \frac{1}{\tau_{gr}}\right] \delta\nu_{gc} - \left[ \frac{1}{\tau_{rc}} + \frac{I_g}{I_r+I_g}\frac{1}{\tau_{gr}}\right] \delta\nu_{rc}, \label{eq:d_dnu_rc}
    \end{align}
    where $\tau_{ij} = (\alpha_{ij} + \alpha_{ji})^{-1}$ are characteristic time-scale of the interaction between i-th and j-th components,
    \begin{equation}
      b = \left( \frac{\partial A}{\partial\nu_{gs}}\right)_{\nu_{gs}^{eq}}.
    \end{equation}
    }
    Note that these equations are obtained under the assumption that $|\delta\nu_{gs}| \ll \omega_\mathrm{crit}$.
    Knowing  $\delta\nu_{gc}(t)$, one can calculate the departure of the crust rotation from the steady-state spin-down:
    \begin{equation}
      \label{eq:DOmega}
      \Delta\Omega_c(t) = \Omega_c(t)  + |\dot{\Omega}| t = \frac{I_g}{I_g+I_c}\frac{1}{\tau_{gc}} \int_0^t \delta\nu_{gc} dt.
    \end{equation}
    Equations \eqref{eq:d_dnu_gc}-\eqref{eq:d_dnu_rc} can be solved numerically for each particular case. 
    However, in present work, we only make some estimations.
    
    \begin{figure}
          \center
          \vspace{-1cm}
          \includegraphics[width=0.7\textwidth, trim= 0cm 0cm 0.6cm 0.cm, clip = true]{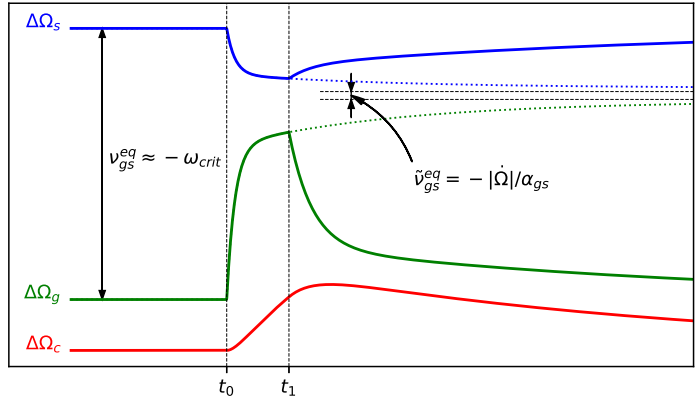}
          \caption{\label{fig:glitch}The sketch of the evolution  of the corrections to the components steady-state spin-down $\Delta\Omega_i = \Omega_i + |\dot{\Omega}|t$ during the glitch. }
          \end{figure}       
    Let us assume that just before the glitch all the corrections are equal to zero. 
    At $t=t_0$, due to some fluctuation or an external influence, the superfluid vortices become unpinned. 
    It can be formally described by putting the activation energy $A$ 
    to be identically equal to zero in all expressions.
    According to equation \eqref{eq:nu_gs_eq}, it leads to the change of steady-state lag ${\nu}_{gs}^{eq}$ which becomes equal to $\tilde{\nu}_{gs}^{eq} = -{|\dot{\Omega}|}/{\alpha_{gs}}$ instead of ${\nu}_{gs}^{eq}\approx - \omega_\mathrm{crit}$.
    Thus,  at $t=t_0$, we have  $\delta \nu_{gs}+ \tilde{\nu}_{gs}^\mathrm{eq} = \nu_{gs}^\mathrm{eq}$.
    The steady-state values of $\nu_{gc}$ and $\nu_{rc}$ do not change after the unpinning event.
    As a result, at $t=t_0$, 
    the system starts to evolve with the following initial conditions: $\delta \nu_{gs}(t_0) = - \omega_\mathrm{crit} + {|\dot{\Omega}|}/{\alpha_{gs}}, \delta \nu_{gc}(t_0)=\delta \nu_{rc}(t_0)=0$. 
    During this stage, the superfluid injects the stored angular momentum into the g-component reducing $\nu_{gs}$ by absolute value. 
    At some moment $t=t_1$ the vortices repin and the system starts to evolve backwards.
    The stages of the glitch are shown in Figure \ref{fig:glitch} in terms of the component angular velocities.

    We have assumed that the r-component is responsible for long-term glitch recovery \cite{GB2019}, i.e. $\tau_{gr}, \tau_{rc} \sim (1-100)$ days. Hence, during the glitch spin-up, the r-component can be ignored in most cases. The r-component angular velocity is not shown in Figure \ref{fig:glitch}.
    The rate of the spin-up is determined by  the larger of time-scales $\tau_{gc}$ and $\tau_{gs}$.  
    The first is difficult to calculate because it depends on the poorly known properties of the neutron star matter in the vicinity if the crust-core region as well as the geometry of the g-component \cite{GB2019}.
    The second is determined by the electron scattering on the superfluid vortices \cite{AlparLangerSauls1984}
    and the corresponding coefficient could be estimated as $\alpha_{gs} \sim 5 \times 10^{-2} (q P)^{-1}$ \cite{BGT2013},
     where $P$ is the period of pulsar rotation and $q$ is the coefficient determined by the proton effective mass. Coefficient $q$ is estimated as $\sim 10 - 200$ and it decreases depthward the star \cite{SideryAlpar2009}.
     Since the g-component is rather located near the crust-core interface, we choose $q=100$.
     In this case,  $\tau_{gs} = I_g/(I_g+I_s) \{66\mbox{ s (Crab)},  \ 180\mbox{ s (Vela)}, \ 830\mbox{ s (B1828-11)}\}$. 
    If  the whole g-component superfluid is unpinned during the glitch, then $I_g/(I_g+I_s) \sim 0.1$.
    For the glitch of 2017 year Shaw et al. \cite{ShawEtAl2018} placed an upper limit $\tau_\mathrm{glitch}<6$ hr for the unresolved component of the spin-up, while the delayed spin-up occurred over a time-scale of $\sim$1.7 d.
    These time intervals are much larger than estimated value $\tau_{gs}$. It could indicate that the Crab pulsar contains the  magnetically decoupled g-component. The analysis of equations \eqref{eq:d_dnu_gc}-\eqref{eq:DOmega} allows us to relate the  gs-correction to the steady state lag required to be induced to provide the observed glitch size: 
    \begin{equation}
      \delta \nu_{gs}  \lesssim \frac{I_g+I_s}{I_s}\frac{I_g+I_c}{I_g} \Delta \Omega_\mathrm{glitch}
    \end{equation}
    Taking into account that $I_c/I_g \lesssim 10-100$ \cite{GB2019}, 
    we obtain $\delta \nu_{gs} \approx 10^{-3} - 10^{-2} \mbox{s}^{-1}$. On the other hand, the steady-state lag  $\nu_{gs}^{eq} \approx \omega_\mathrm{crit} \approx 10^{-1}\mbox{s}^{-1}$ \cite{Link2014}. 
    Thus, the angular momentum stored by superfluid is enough even for the largest Crab glitch. 
    
    Considering the Vela pulsar, the spin-up of 2016 year glitch lasted only a few seconds \cite{PalfreymanEtAl2018}. According to our estimations $\tau_{gs}  = 18$ s. Hence, the possibility of the glitch originating from the magnetically decoupled core region is more sensitive to the model parameters.  
    For instance,  the g-component could include more deeper regions  such that coefficient $q$ is effectively smaller.
    
    For B1828-11 to date only one small glitch was registered\footnote{http://www.jb.man.ac.uk/pulsar/glitches.html} \cite{EspinozaLyneStappersKramer2011}. Unfortunately, there are no known time-scales for this glitch.

    Our simple estimations show that the 2017 year Crab glitch could originate from the magnetically decoupled core region. For the 2016 year Vela glitch whose spin-up time is restricted much more stronger the results is more controversial.
    Strictly speaking, the proposed model does not have to describe all glitching pulsars.  In a part of pulsar population the g- and the c-component could be connected rigidly. This population, however, can not precess with long periods.


  \section*{References}
    \bibliography{mn-jour,paper}   
    
\end{document}